\documentclass[fleqn,usenatbib]{mnras}
\DeclareRobustCommand{\VAN}[3]{#2}
\let\VANthebibliography\thebibliography
\def\thebibliography{\DeclareRobustCommand{\VAN}[3]{##3}\VANthebibliography}
\usepackage{graphicx}	
\usepackage{amsmath}	
\usepackage{amssymb}	
\usepackage{colortbl}
\usepackage{float}
\usepackage{threeparttable}
\usepackage[FIGTOPCAP,nooneline,scriptsize]{subfigure}
\definecolor{mygray}{gray}{.9}

\title[Simultaneous observation of Swift~J1818.0$-$1607]{Simultaneous 2.25/8.60 GHz observations of the newly discovered magnetar - Swift~J1818.0$-$1607}

\author[Zhi-Peng Huang et al.]{
Zhi-Peng Huang,$^{1,2,3}$
Zhen Yan,$^{1,2,4}$\thanks{E-mail: yanzhen@shao.ac.cn}
Zhi-Qiang Shen,$^{1,2,3,4}$
Hao Tong,$^{5}$
Lin Lin,$^{6}$
Jian-Ping Yuan,$^{4,7}$\newauthor
Jie Liu,$^{1,2,3}$
Ru-Shuang Zhao,$^{8,9}$
Ming-Yu Ge,$^{10}$
and Rui Wang$^{1,2}$
\\
$^{1}$Shanghai Astronomical Observatory, CAS, Shanghai 200030, China\\
$^{2}$University of Chinese Academy of Sciences, Beijing 100049, China\\
$^{3}$School of Physical Science and Technology, ShanghaiTech University, Shanghai 201210, China\\
$^{4}$Key Laboratory of Radio Astronomy, CAS, 10 Yuanhua Road, Nanjing, JiangSu 210033, China\\
$^{5}$School of Physics and Materials Science, Guangzhou University, Guangzhou 510006, China\\
$^{6}$Department of Astronomy, Beijing Normal University, Beijing 100875, China\\
$^{7}$Xinjiang Astronomical Observatory, CAS, Xinjiang 830011, China\\
$^{8}$School of Physics and Electronic Science, Guizhou Normal University, Guiyang 550001, China\\
$^{9}$Guizhou Provincial Key Laboratory of Radio Astronomy and Data Processing, Guizhou Normal University, Guiyang 550001, China\\
$^{10}$Key Laboratory of Particle Astrophysics, Institute of High Energy Physics, CAS, 19B Yuquan Road, Beijing 100049, China
}

\date{Accepted 2021 May 3. Received 2021 May 3; in original form 2020 October 27}

\pubyear{2021}

\begin{document}
\label{firstpage}
\pagerange{\pageref{firstpage}--\pageref{lastpage}}
\maketitle

\begin{abstract}
Swift~J1818.0$-$1607 discovered in early 2020 is not only the fifth magnetar known with periodic radio pulsations but also the fastest rotating one. Simultaneous 2.25/8.60~GHz observations of Swift~J1818.0$-$1607 were carried out with  Shanghai Tian Ma Radio Telescope (TMRT) from MJD 58936 to 59092. The spin-frequency $\nu$ and first-order derivative $\dot{\nu}$ of this magnetar were obtained with piecewise fitting method because of its instable timing properties. We found that the amplitude of short-term $\dot{\nu}$ fluctuations decreased with time, and the long-term declining trend of $\nu$ discovered previously continued in our observations. The best fitted long-term $\dot{\nu}$ were about $ -2.25\times10^{-11} s^{-2}$ using our observation data spanning 156 days. The derived characteristic age was about 522~yr, supporting the recent viewpoint that this magnetar may be older than initially thought shortly after its discovery. The flux density of this magnetar was increased at both 2.25 and 8.60 GHz during our observations, and its radio spectrum became flatter at the same time. We also detected bright-quiet type emission mode switching in Swift~J1818.0$-$1607.
\end{abstract}

\begin{keywords}
stars: neutron --- stars: magnetars --- pulsars: individual: Swift~J1818.0$-$1607
\end{keywords}



\section{Introduction} \label{sec:intro}

Magnetars are a special class of young neutron stars with extremely strong magnetic fields ($10^{13}-10^{15}$~G). They sometimes show dramatic variabilities across the electromagnetic spectrum, especially at high energy band, such as giant flares \citep{mer2008}. The decay of enormous internal magnetic fields of magnetars is thought to power these emissions, as they are too bright to be powered by the loss of rotational energy or the accretion power \citep{dt1992}. By the end of 2020, there were 31 magnetars and candidates discovered \citep{olk14,cbi2021}\footnote{\url{http://www.physics.mcgill.ca/~pulsar/magnetar/main.html}\label{web}}. Compared with other neutron stars, magnetars generally rotate slower and spin down faster because of their stronger magnetic braking effects \citep{txs2013,gpw2019}. Though magnetars show dramatic outbursts at high energy, only 6 magnetars had been detected with radio pulse radiation until the end of 2020 \citep{olk14}\textsuperscript{\ref{web}}. Typically, magnetar shows variable integrated profiles on timescales from hours to days and relatively flat spectra with typical spectral indices ($\alpha$) greater than $-$0.8 at radio band \citep{crp2007,tek2015,tde2017}, which is different from normal pulsars that usually show stable integrated profiles and steeper radio spectra with typical $\alpha$ around $-$1.6 \citep{jsk2018}.

Swift~J1818.0$-$1607 was firstly captured as an X-ray burst by the Swift Burst Alert Telescope on  March 12th 2020. It was confirmed to be a magnetar with a spin period ($P$) of 1.36~s by the follow-up observations of Neutron star Interior Composition Explorer \citep{esy2020}, making it the magnetar with the shortest $P$ known so far. Furthermore, it was shown to be the 5th radio magnetar known, as its radio pulsations were detected by Effelsberg telescope \citep{kdk2020}. Shortly after the X-ray burst, its integrated profile could only be detected below 8.0~GHz, though high frequency observations were arranged \citep{rsl2020,lsj2020,lbj2020,rbc2020,mpp2020b,ccc2020}. About 27 days later, it was detected in several high frequencies observations, such as 8.4~GHz \citep{mpp2020a}, 22~GHz \citep{lkc2020}, 31.9~GHz \citep{pmp2020}, 86 and 154~GHz \citep{tlc2020}. Its spectrum got flatter with $\alpha>-0.97$. In most cases, Swift~J1818.0$-$1607 showed integrated profile with single peak. Secondary peak component on the right of main peak was sometimes detected at 3.8~GHz \citep{lsj2020}, 1.37 and 2.55~GHz \citep{ccc2020}. Furthermore, \cite{ljs2020} detected integrated profile with triple peaks at 2.368~GHz using Parkes Ultra-Wideband receiver.

As more timing observation data obtained, it was discovered that its rotation frequency derivative $\dot\nu$ showed drastic fluctuations, though there was a long-term declining trend of $\nu$. Besides the glitch detected in Swift~J1818.0$-$1607 \citep{ccc2020,hbb2020}, it also possibly showed an anti-glitch \citep{hbb2020}.

Long-term simultaneous multi-frequency observations are vital for studying changes in radio emission of magnetars in both frequency and time. Simultaneous 2.25/8.60~GHz observations of Swift~J1818.0$-$1607 were arranged at Shanghai Tian Ma Radio Telescope (TMRT) to do further studies on its radiation and rotational properties. Information about observations is provided in Section~\ref{sec:obser}. The data reduction methods and results are given in Section~\ref{sec:result}. Finally, further discussions are presented in Section~\ref{sec:discussion}.

\section{Observations} \label{sec:obser}
Simultaneous 2.25/8.60~GHz observations were carried out ranging from MJD~58936 (Mar 28th 2020) to MJD~59092 (Aug 31st 2020) with the TMRT by taking advantage of the parallel working capabilities of the 2.25/8.60~GHz dual-frequency receiver and the digital backend system (DIBAS) which consisted of 3 pairs of analog-to-digital converters and Roach-2 boards  \citep{ysw18}. The dual-frequency receiver is a cryogenically cooled, dual-polarization receiver with the frequency coverage of 2.20-2.30~GHz and 8.20-9.00~GHz, respectively. The total bandwidth was divided into channels with a width of 1 MHz (at 2.25~GHz) and 2 MHz (at 8.60~GHz) with the DIBAS to remove the dispersion effects and radio-frequency interferences (RFIs) \citep{ysw15}. The incoherent de-dispersion and online-folding observation mode were used in our observations.  Each rotation period was divided into 1024 pulse phase bins and folded with 30~s sub-integration length. The observation data were written out as the 8-bit PSRFITS format \citep{hvm2004}.

After removing RFIs with the \texttt{pazi} command of the Pulsar Archive analysis software (\texttt{PSRCHIVE}) \citep{hvm2004}, the data was compressed by scrunching all frequency channels and polarizations together with the \texttt{pam} command.

There was no flux calibrator observations arranged before MJD~59039. By assuming the off-pulse root-mean-square (RMS) noise following radiometer equation, we estimated the peak flux density of Swift~J1818.0$-$1607 based on typical system equivalent flux density (SEFD), and the SEFD of the TMRT is about 46~Jy and 48~Jy at the corresponding frequencies \citep{ysw18}. Then, the mean flux density was obtained by dividing the integrated flux density from the on-pulse bins by the total number of phase bins. The uncertainty in mean flux density was obtained from off-pulse RMS noise divided by square root of the number of phase bins. \citep{zwy17}. For the observations later than MJD~59039, we estimated its flux density with both SEFD method and calibrator 3C~295. According to our calculations, these two methods are in agreement with each other within the error bar.

\begin{table}
\scriptsize
\linespread{1.45}
\centering
\caption{Parameters of Swift~J1818.0$-$1607 observations}
\vspace{-0.1cm}
\label{Tab:table1}
\begin{tabular}{c c c c c}
\hline
\hline
\multicolumn{1}{c}{MJD$^a$} &
\multicolumn{1}{c}{Length} &
\multicolumn{1}{c}{$S_{\rm 2.25}/S_{\rm 8.60}$$^b$} &
\multicolumn{1}{c}{$\alpha$$^c$} &
\multicolumn{1}{c}{Shape$^d$}
\\
\multicolumn{1}{c}{\footnotesize{--}} &
\multicolumn{1}{c}{\footnotesize{(h)}} &
\multicolumn{1}{c}{\footnotesize{(mJy)}} &
\multicolumn{1}{c}{\footnotesize{--}} &
\multicolumn{1}{c}{\footnotesize{--}}
\\
\hline
58936.80 & 5.7 & $1.31\pm0.07$/$0.11\pm0.02$ & $-1.84\pm0.14$ &S/T\\
\rowcolor{mygray}
58937.79 & 6.0 & $1.26\pm0.06$/receiver fault & -- &S/-\\
\rowcolor{mygray}
58938.79 & 6.0 &  $1.35\pm0.07$/$0.12\pm0.02$ & $-1.78\pm0.14$ &S/T\\
58941.79 & 4.7 & $1.04\pm0.06$/$0.14\pm0.02$ & $-1.48\pm0.13$  &S/T\\
\rowcolor{mygray}
58944.79 & 5.3 & $1.00\pm0.06$/$0.11\pm0.02$ & $-1.63\pm0.13$ &S/T\\
59015.66 & 0.8 & $0.86\pm0.13$/$0.48\pm0.06$ & $-0.43\pm0.14$ &S/S\\
59020.53 & 1.9 & $0.61\pm0.10$/$0.27\pm0.04$ & $-0.60\pm0.16$ &S/S\\
\rowcolor{mygray}
59029.52 & 2.0 & $0.45\pm0.09$/$0.32\pm0.03$ & $-0.24\pm0.17$ &D/S\\
\rowcolor{mygray}
59030.49 & 2.0 & $0.42\pm0.09$/$0.59\pm0.03$ & $+0.26\pm0.16$ &D/S\\
59036.47 & 1.0 & receiver fault/$0.33\pm0.05$ & -- &-/S\\
\rowcolor{mygray}
59039.48 & 0.5 & $0.77\pm0.18$/$0.78\pm0.07$ & $+0.01\pm0.19$ &D/S\\
\rowcolor{mygray}
59040.47 & 0.5 & receiver fault/$0.51\pm0.09$ & -- &-/S\\
\rowcolor{mygray}
59041.45 & 0.5 & receiver fault/$0.30\pm0.07$ & -- &-/S\\
59049.49 & 0.5 & $1.19\pm0.21$/$0.80\pm0.08$ & $-0.30\pm0.15$ &D/S\\
59050.50 & 0.4 & $1.21\pm0.18$/$1.66\pm0.07$ & $+0.24\pm0.12$ &D/S\\
\rowcolor{mygray}
59051.47 & 2.0 & $1.58\pm0.09$/$3.76\pm0.03$ & $+0.65\pm0.05$ &D/S\\
\rowcolor{mygray}
59052.47 & 2.0 & $1.02\pm0.09$/$1.61\pm0.03$ & $+0.34\pm0.07$ &D/S\\
59070.49 & 4.1 & $1.77\pm0.07$/$1.57\pm0.02$ & $-0.09\pm0.03$ &D/S\\
59073.48 & 5.6 & $2.08\pm0.06$/$1.28\pm0.02$ & $-0.36\pm0.02$ &D/S\\
\rowcolor{mygray}
59074.45 & 6.0 & $1.80\pm0.07$/$1.72\pm0.02$ & $-0.03\pm0.03$ &D/S\\
59075.44 & 5.9 & $1.92\pm0.06$/$1.64\pm0.02$ & $-0.12\pm0.02$ &D/S\\
\rowcolor{mygray}
59076.42 & 6.0 & $1.71\pm0.06$/$1.84\pm0.02$ & $+0.05\pm0.03$ &D/S\\
59077.45 & 5.9 & $1.79\pm0.06$/$1.32\pm0.02$ & $-0.23\pm0.03$ &D/S\\
\rowcolor{mygray}
59078.45 & 5.8 & $1.18\pm0.06$/$1.67\pm0.02$ & $+0.26\pm0.04$ &D/S\\
59079.44 & 5.9 & $3.02\pm0.06$/$2.47\pm0.02$ & $-0.15\pm0.01$ &D/S\\
\rowcolor{mygray}
59080.40 & 6.0 & $2.65\pm0.05$/$2.45\pm0.02$ & $-0.06\pm0.02$ &D/S\\
59081.43 & 3.9 & receiver fault/$0.68\pm0.02$ & -- &-/S\\
\rowcolor{mygray}
59082.42 & 4.3 & receiver fault/$2.37\pm0.02$ & -- &-/D\\
59086.48 & 3.4 & $2.62\pm0.07$/$1.20\pm0.02$ & $-0.58\pm0.02$ &D/D\\
\rowcolor{mygray}
59088.40 & 4.5 & $1.65\pm0.06$/$1.38\pm0.02$ & $-0.13\pm0.03$ &D/D\\
59090.53 & 2.2 & $3.04\pm0.09$/$1.30\pm0.03$ & $-0.63\pm0.03$ &T/D\\
59092.47 & 3.5 & $3.01\pm0.07$/$1.86\pm0.03$ & $-0.36\pm0.02$ &D/D\\
\hline
\multicolumn{5}{l}{\scriptsize$^a$ Start time.}\\
\multicolumn{5}{l}{\scriptsize$^b$ Mean flux density at 2.25/8.60~GHz.}\\
\multicolumn{5}{l}{\scriptsize$^c$ Spectral index.}\\
\multicolumn{5}{l}{\scriptsize$^d$ Integrated profile morphology: Single (S), Double (D) and Triple (T) peaks.}
\end{tabular}
\end{table}

\section{Results} \label{sec:result}
\subsection{Evolution of spin frequency and DM}
\begin{figure}
\centering
\includegraphics[width=0.46\textwidth]{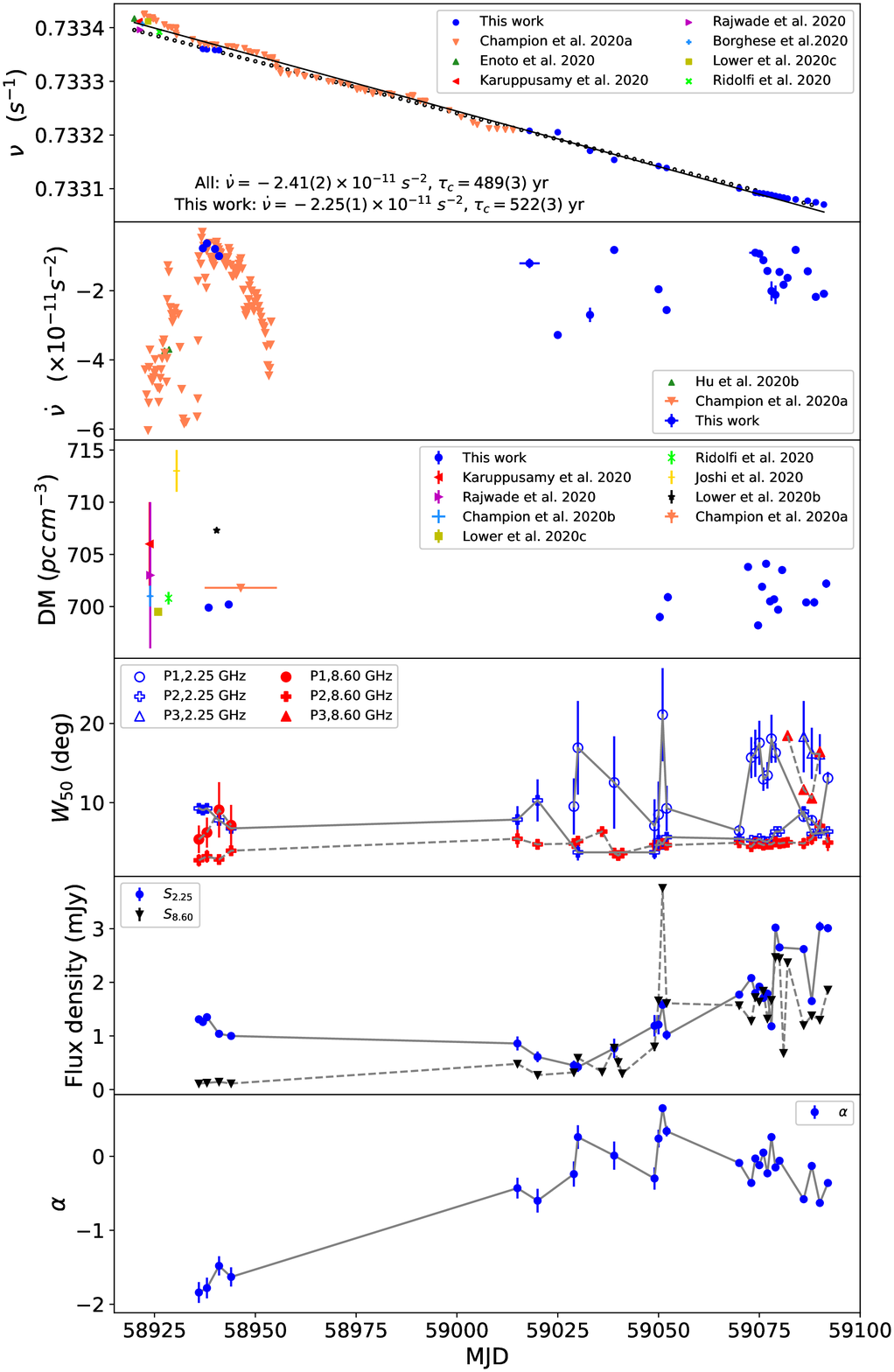}
\caption{The plots about $\nu$, $\dot\nu$, DM, $W_{\rm 50}$, $S_{\rm 2.25}$, $S_{\rm 8.60}$ and $\alpha$ of Swift~J1818.0$-$1607 changing with the time. The best fitted long-term $\nu$ for our observation data and all observation data are shown with solid and dot line respectively in top panel.}
\label{fig:Figure1}
\end{figure}
Both the integrated profile and rotational properties of Swift~J1818.0$-$1607 were found to change irregularly during our observations, which makes constructing a phase$-$coherent timing solution very difficult. Thus, we divided our observations into segments and carried out timing analysis to each of them separately, similar to the strategy presented in previous work such as \cite{lbb2012}. The template profiles of 2.25/8.60~GHz were respectively obtained based on the integrated profile of 2.25/8.60~GHz with typical SNRs $>$10.28 in the middle of each segment. Then, the time-of-arrivals (ToAs) of pulses was obtained by cross-correlating integrated profiles with the standard template of corresponding frequencies with \texttt{pat} command. Next, we combined the ToA sets from 2.25/8.60~GHz after removing the delay of the receiver system, and carried out timing analysis with \texttt{TEMPO2} software package \citep{hem2006} to measure the spin properties of the magnetar. The length of each segment lasted until there were obvious changes in the integrated profiles or the error of fitted $\dot\nu$ exceeded $2.70\times 10^{-12}$~$\rm s^{-2}$ which was about 10\% of mean $\dot\nu$ \citep{ccc2020}. In Table~\ref{Tab:table1}, two adjacent segments are distinguished with different background colors. The top three panels of Fig.~\ref{fig:Figure1} show the fitted $\nu$, $\dot\nu$ and DM changing with the time along with some previous published results \citep{ccc2020,cdj2020,esy2020,bzr2020,rsl2020,lsj2020,lbj2020,ls2020,rbc2020,hsr2020,jb2020}. The data points from \cite{ccc2020}'s work were captured from the published plots with the open-source software \texttt{engauge digitizer} \footnote{http://markummitchell.github.io/engauge-digitizer/}.

It's clear that the long-term decline trend of $\nu$ previously reported by \cite{ccc2020} and \cite{hbb2020} went on in our observations, though $\dot\nu$ still changes erratically on short time-scale. We did a linear fit to the $\nu$ measurements with the least square method base on only ours and all observation results shown in the top panel of Fig.\ref{fig:Figure1}. The best-fitted $\dot{\nu}$ was about $-2.25(1)\times10^{-11} s^{-2}$ and $-2.41(2)\times10^{-11} s^{-2}$, respectively. \cite{ccc2020} also reported a similar fitted $\dot{\nu}$ of  $-2.37\times10^{-11} s^{-2}$ using the data spanning nearly 100 days. Combined with all observations, the $\dot{\nu}$ showed continuous oscillations instead of random variations after the discovery burst. It also was noticed that the amplitude of the oscillation seems to decrease in time. The DM values obtained with TMRT are consistent with previous values. The DM range is between 699 and 705 $pc\enspace cm^{-3}$ except for three observations before MJD~59040 \citep{kdk2020,jb2020,lsj2020}.

\subsection{Integrated profiles}
\begin{figure}
\begin{center}
\includegraphics[width=0.46\textwidth]{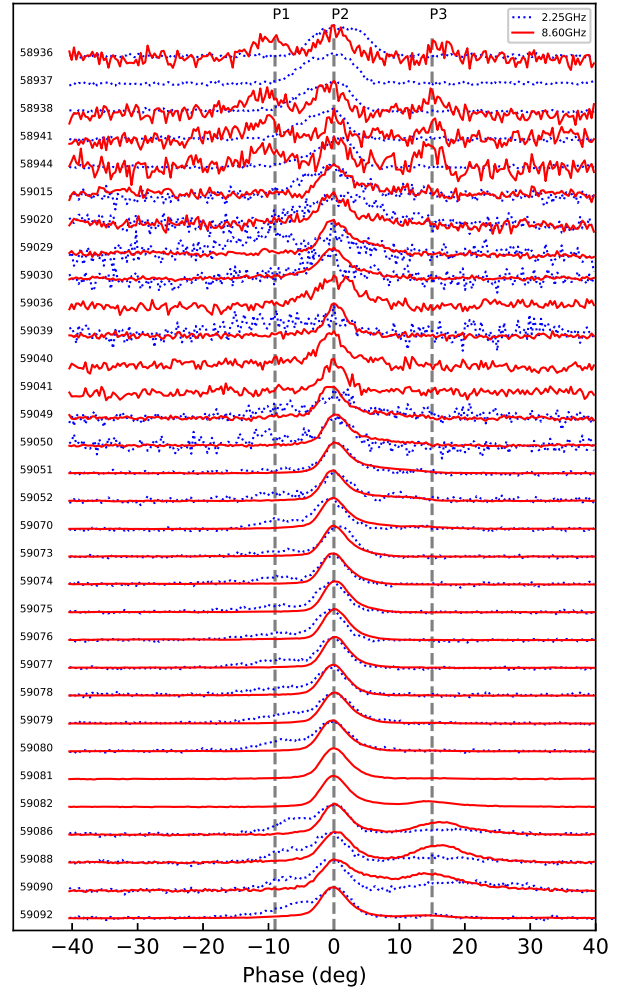}
\end{center}
\vspace{-0.6cm}
\caption{Normalized integrated profiles for Swift~J1818.0$-$1607 obtained  at 2.25~GHz (blue dotted line) and 8.60~GHz (red solid line). The typical peaks of integrated profiles are connected with vertical grey dotted lines.} \label{fig:Figure2}
\end{figure}
Integrated profiles of pulsars provide important information of the structure of the radiation beam. With updated ephemeris, the integrated profiles for each epoch of 2.25/8.60~GHz observations were obtained and showed chronologically in Fig.~\ref{fig:Figure2}. For easier comparison, they are normalized, aligned with respect to the peak of the profile and centered at the zero rotation phase. Three phase areas which are usually giving out radiation are labeled with $P1$, $P2$ and $P3$ from left to right. From the plots in Fig.~\ref{fig:Figure2}, we can see that the integrated profiles of Swift~J1818.0$-$1607 can keep stable in several days at both 2.25 and 8.60~GHz. The integrated profile morphology in each epoch of our observations is also listed in the last column of Table~\ref{Tab:table1}. In most cases, its integrated profile showed single peak-structure at 8.60~GHz, whereas it usually showed dual-peak structure at 2.25~GHz.  It is noticed that the integrated profile shapes at 2.25/8.60~GHz only showed synchronized changes during MJD~59086-59092 on the $P3$ component. We fitted three Gaussian components to each profile to obtain the width at 50\% of peak ($W_{\rm 50}$) of each component. The curves showing how $W_{\rm 50}$ of each component changed with the time are shown in the fourth panel of Fig.~\ref{fig:Figure1}. The $W_{\rm 50}$ of $P2$ at 8.60~GHz is generally narrower than that at 2.25~GHz except on MJDs from 59030 to 59051. Normally, the $W_{\rm 50}$ of two other components is wider than the $W_{\rm 50}$ of $P2$ both at 2.25 and 8.60~GHz.

\subsection{Flux densities and spectral index}
The estimated flux density at 2.25/8.60~GHz ($S_{\rm 2.25}$ and $S_{\rm 8.60}$) are appended in the 3rd column of Table~\ref{Tab:table1} and also shown in 5th panel of Fig.~\ref{fig:Figure1}. Though $S_{\rm 2.25}$ decreased before MJD~58944 shortly after discovery X-ray burst, there was a steady rising trend in both $S_{\rm 2.25}$ and $S_{\rm 8.60}$ after MJD~59036. For each epoch, we fitted a power-low model ($S_{\nu} \propto \nu ^{\alpha}$) to the measured flux densities, and showed the spectral index $\alpha$ in the 4th column of Table~\ref{Tab:table1} and also gave related plots in the bottom panel of Fig.~\ref{fig:Figure1}. It can be seen that magnetar changed from steep spectrum with $\alpha < -1.48$ (MJD~58936-58944) to flat spectrum with $\alpha > -0.63$ (MJD~59015-59092). Our $\alpha$ results are consistent with previous values obtained below 32~GHz \citep{mpp2020a,mpp2020b,pmp2020,lkc2020}. Compared the $\alpha=+0.5$ obtained by us on MJD~59076, \cite{tlc2020} reported $\alpha=-1.4$ with observations from 86 to 154~GHz on same day. So, we guess there should be at least a spectra turnover at high frequency.

\subsection{Mode switching}
\begin{figure}
\begin{center}
\includegraphics[width=0.49\textwidth]{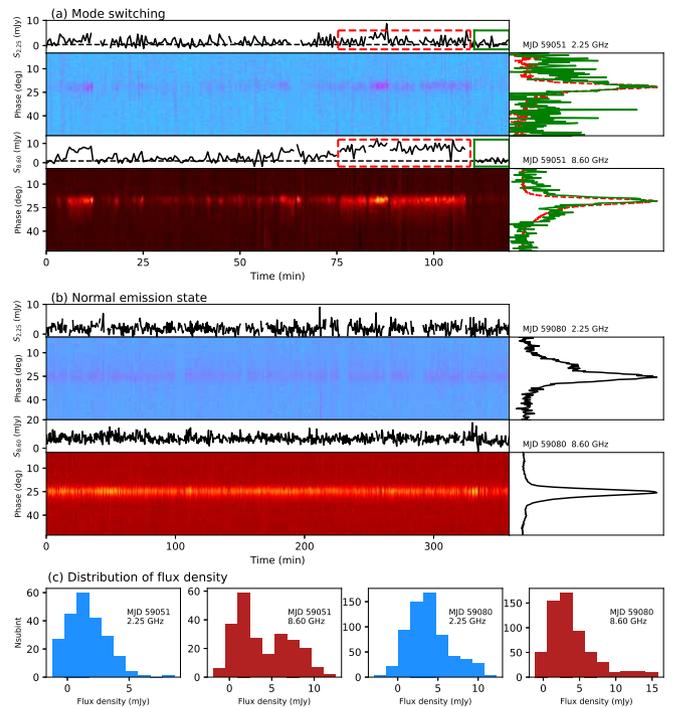}
\vspace{-0.5cm}
\caption{Plots for mode switching of Swift~J1818.0$-$1607 on MJD~$59051$ and normal emission state on MJD~$59080$: phase-time waterfall plots (`jet' for 2.25~GHz;  `hot' for 8.60~GHz); mean flux density change with time (above subpanel of subfigure~a and b); integrated profiles (right subpanel of subfigure~a and b); histograms of mean flux density distribution  (subfigure~c)}\label{fig:Figure3}
\end{center}
\end{figure}

Besides mild changes, Swift~J1818.0$-$1607 showed obvious short time-scale flux density variations on MJDs~59039, 59050 and 59051. It seems to switch between quasi-stable bright (B) and quiet (Q) radiation state. Diagnostic plots for its B-Q radiation state are shown in Fig.~\ref{fig:Figure3} along with relating plots for its normal radiation state. By comparison, it is a little easier to judge whether it is a mode switching or not using data from MJD~59051 than the other two data because of its relatively higher average flux density. Subfigure~(a) of Fig.~\ref{fig:Figure3} shows the flux density changes at 2.25 and 8.60~GHz on MJD~59051 with the `hot' and `jet' style phase-time plots, respectively. Its mean flux density changes with the time are shown in the subpanel above each waterfall plot with the black solid curve. To avoid being contaminated by adjacent data of different states, a typically B and Q state that lasted relatively longer were selected and labeled with red dash and green solid square, respectively. The black dash line in the same panel stands for 10\% of the average flux of B state labeled with the red dash square. It can be seen that this magnetar switched between B and Q state synchronously at 2.25/8.60~GHz. The integrated profiles obtained the Q state data in green solid squares with same color are shown in right subpanels. Meanwhile, the integrated profiles of its B state obtained same style curve data in the red dash squares are shown with the red line in the right subpanel of Fig.\ref{fig:Figure3}. We can see the difference in shapes of red and green integrated profiles, especially for 8.60~GHz observation. In Q state, its integrated profile manifested as an additional bump trailing the main peak at 8.60~GHz, but the bump vanished in B state. Furthermore, it can be clearly seen in subfigure~(c) of Fig.\ref{fig:Figure3} flux density distribution for 8.60~GHz observation on MJD~59051 shows a double-peak structure. Therefore, we think the magnetar exhibited mode switching on MJD 59051 (plots for the B-Q mode switching on MJD~59039 and 59050 will be given in supporting materials).

\section{Discussions} \label{sec:discussion}
Swift~J1818.0$-$1607 showed irregular timing properties. As shown in the 2nd panel of Fig.~\ref{fig:Figure1}, it seems that the $\dot\nu$ of Swift~J1818.0$-$1607 oscillated but not a stochastic process. The amplitude of $\dot\nu$ oscillations seemed to get smaller as time went. Similar timing phenomena were observed in the magnetar 1E~1048.1$-$5937 and 17 ordinary pulsars but with timescales of years \citep{lhk2010,ask2020}. These $\dot\nu$ fluctuations may be caused by certain types of unstable torque that may be affected by stronger magnetic braking \citep{ccr2007,lbb2012,tx2013}.

Assuming an idealized magnetic dipole model in a vacuum (braking index $n = 3$) and much faster spin at birth, its characteristic age $\tau_{\rm c}$ was initially reported as 265, 310 and 240~yr \citep{cdj2020, hsr2020, erb2020}, making it be the youngest magnetar known. The $\dot\nu$ of Swift~J1818.0$-$1607 fluctuated more than 12 times from MJD~58930 to 58940 \citep{ccc2020}. Though the $\dot\nu$ showed obvious fluctuations in short time-scale, there was obvious linear decrease trend of $\nu$ in long time-scale. Using much longer observation data spanning about 100 days, \cite{hbb2020} and \cite{ccc2020} fitted the average $\dot\nu$ then got relatively older $\tau_{\rm c}$ age of 470 and 500~yr, respectively. We found the long-term linear decrease of $\nu$ went on in our observations.  Base on the best fitted $\dot\nu$ obtained with only ours and all the data in Fig.~\ref{fig:Figure1}, the $\tau_{\rm c}$ inferred with the above assumptions should be 522 and 489~yr, respectively. So, the $\tau_{\rm c}$ of Swift~J1818.0$-$1607 is not as young as 300~yr.

Normally, radio-loud magnetars show a flat spectrum typically with $\alpha>-0.8$ \citep{efk2013,tde2017}. In our observations of Swift~J1818.0$-$1607, there was a long-term flux density increase trend at both 2.25 and 8.60~GHz. The $\alpha$ changed from less than $-$1.48 to greater than $-$0.63 as time went on. Judging from plots in Fig.~\ref{fig:Figure1}, we think its steep spectrum that lasted tens of days was the main reason for no successful detection of integrated profile above 8.0~GHz \citep{gps2020,ls2020,mpp2020a}. The steep radio spectrum was temporarily detected on some magnetars in time-scale of several hours \citep{ljk2008,pmp2019}. A steep magnetar radio spectrum that lasted for tens of days has so far only been observed in J1745$-$2900, soon after its early outburst \citep{ppe2015}.

Magnetars usually change their integrated profile shapes dramatically on timescales from hours to days \citep{ppe2015}. Besides integrated profile with single-peak and dual-peak structure usually detected at frequencies from 0.8 to 154~GHz, we also detected triple-peaked structure at 2.25~GHz near MJD 59087 as Parkes \citep{ljs2020}. Though two new components showed at 8.60~GHz, they were not seen at 2.25~GHz from MJD 58936 to 58944 in our observations. \cite{hje2015} also detected phenomena in which new component dominated profile at high frequency in the Crab Pulsar.

\cite{ljs2020} reported two types of mode switching of Swift~J1818.0$-$1607 at 0.7-4.0~GHz. Swift~J1818.0$-$1607 was quasi-periodically switching between B-mode and Q-mode on MJD~59009, and varying between two longitudinally distinct modes (P- and M-mode) on MJD~59047. Swift~J1818.0$-$1607 switched between B-mode and Q-mode synchronously at 2.25/8.60~GHz in 3 days of our observation. Though the mode switching detected by us took place only 3 days later than `P-M' mode switching on MJD~59047, its flux density fluctuation was more similar to `B-Q' mode switching on MJD~59009. But, the right bump of integrated profile appeared in the Q-mode of our observation at 8.60~GHz was not so strong as \cite{ljs2020}'s result in which it was brighter than the left peak component.

\section*{Acknowledgments}
This work was supported in part by  the Natural Science Foundation of Shanghai (Grant No.~20ZR1467600),  National Natural Science Foundation of China (Grant Nos. U2031119, U1631122 and 11633007), Strategic Priority Research Program of the CAS (XDB23010200) and National Key R\&D Program of China (2018YFA0404602).

\section*{Data Availability}
The data underlying this article will be shared on reasonable request to the corresponding author.


\label{lastpage}
\end{document}